# Prospects for Probing Feedback from the First Black Holes and Stars During Reionization

Jack O. Burns[a,b] for the LUNAR[c] Consortium

[a]Center for Astrophysics & Space Astronomy, 389 UCB, University of Colorado, Boulder, CO 80309
[b]NASA Lunar Science Institute, NASA Ames Research Center, Moffett Field, CA
[c]Lunar University Network for Astrophysics Research (LUNAR), http://lunar.colorado.edu

**Abstract.** The feasibility of making highly redshifted HI 21-cm (rest frame) measurements from an early epoch of the Universe between the Dark Ages and Reionization (i.e., z>6 and ν<200 MHz) to probe the effects of feedback from the first stars and quasars is assessed in this paper. It may be possible to determine the distribution of hydrogen through the Universe and to constrain the birth of the first stars and black holes via HI tomography. Such observations may also place limits on the properties of Inflation and any exotic heating mechanisms before the first star formation begins (e.g., dark matter decay). The global (all-sky) HI signal after Recombination has distinct features at different frequencies between 30 and 200 MHz that changes as the relative balance between the CMB and spin temperatures changes due to the expansion of the Universe and the ignition of stars and/or black holes. A technology roadmap to approach these observations beginning with ground-based arrays and ending with a low frequency radio array on the lunar farside is described.

**Keywords:** cosmology, 21-cm measurements, dark ages, epoch of reionization, first stars, first AGNs.
**PACS:** 90.

## THE FIRST EPOCH OF FEEDBACK

Presentations at this conference have made it clear that feedback from stars and central AGN engines are fundamental to understanding the nature of galaxies and galaxy clusters. Feedback played an even more crucial role in an earlier epoch of the Universe as the IGM transformed from a neutral state to the totally ionized plasma that we observe today. How did this transformation occur? How did the first stars and black holes which drove the heating of the IGM evolve (and relate) to the feedback in galaxies seen today via the superb X-ray and radio images?

In recent years, we have begun to realize that probing this first epoch of feedback between the end of the so-called Dark Ages and Reionization may be feasible using highly redshifted 21-cm measurements at corresponding low radio frequencies (<200 MHz) [1]. As shown schematically on the left side in Figure 1, the Universe first becomes transparent during Recombination (z ≈ 1100), produces the CMB, and commences the Dark Ages (a universe with a baryonic content of primarily HI). During this epoch, overdensities of dark matter gravitationally collapse pulling with them baryons (HI). These overdensities eventually produce the first stars and cosmic twilight of the Universe probably at z~30. The precise nature of these first stars (Pop III stars with a Salpeter-like or supermassive IMF?) is unclear. They go through their stellar lifecycle quickly exploding as Type II supernovae, creating large HII regions or bubbles of ionized gas. As more of these first stars form, they produce additional bubbles which begin to merge and form a fully ionized intergalactic medium. During this period, the first black holes and quasars are thought to occur generating intense X-ray emission further driving the reionization. The epoch of reionization likely ends at z>6-10. The details of this schematic picture remain an active area of research and without any observational confirmation.

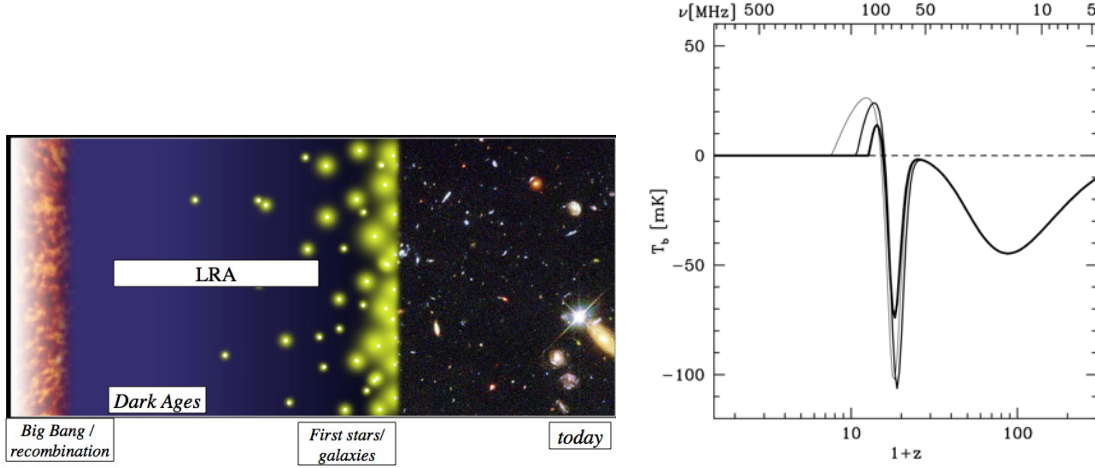

**FIGURE 1.** (Left) Schematic illustration of the evolution of the Universe (Djorgovski et al. & Digital Media Center, Caltech). The first stars and black holes form during the end of the Dark Ages transforming the IGM from a neutral (HI) to an ionized state. LRA is the Lunar Radio Array that would observe highly redshifted 21-cm signals that will constrain the first epoch of feedback. (Right) Evolution of the global (all-sky) redshifted 21-cm brightness temperature as function of redshift (and frequency) for 3 models of the first galaxies [2].

Recently, the global or sky-averaged HI signal expected from the end of the Dark Ages through Reionization (6<z<300) has been modeled [2]. That model spectrum is shown on the right side of Fig. 1. The differential brightness temperature of the redshifted 21-cm line relative to the CMB is given by [3]

$$\delta T_b(z) \approx 23\ (1+\delta)\ x_{HI}\ (1-T_{CMB}/T_S)\ (50\ \Omega_B\ h^2)\ [(0.15\ \Omega_M^{-1}\ h^{-2})\ ((1+z)/10)]^{1/2}\ \text{mK}\quad (1)$$

where $\delta$ is the local matter overdensity, $x_{HI}$ is the neutral fraction of IGM hydrogen, and $T_{CMB}$ is the CMB temperature at z. After recombination, the gas (and spin) temperature tracks the CMB temperature ($T_{CMB}\sim(1+z)$) down to $z \approx 300$ and then declines below it ($T_K\sim(1+z)^2$) producing the right-most absorption trough in Figure 1 corresponding to 5<ν<50 MHz during the Dark Ages. The spin temperature ($T_S$) of the 21-cm transition closely follows $T_K$ through collisional coupling during this time. As the Universe continues to expand and the density decreases, collisional coupling is no longer effective so the spin temperature again tracks the CMB via radiative coupling. Thus, the HI absorption trough closes at $z \approx$ 20-30. Any exotic heating mechanisms such as Dark Matter decay will be evident in the shape and peak of this Dark Ages absorption profile [1].

A second, narrower but deeper, absorption trough is centered at ν ≈ 75 MHz (z = 15-20). This HI absorption is produced after the first stars turn on and begin to flood the IGM with Lyα UV photons. This produces a Lyα resonance with the HI hyperfine transition creating enhanced absorption via the Wouthuysen-Field effect [4]. The shape, width, and center frequency of this trough depend upon the astrophysical details of the first galaxies including the nature of the first stars and X-ray emission from the first black hole-driven AGNs. Thus, this frequency range is a particularly exciting one to probe for details on the first generation of feedback.

As the IGM is heated, most likely by feedback via soft X-rays from the first AGNs, the kinetic and spin temperatures rise above $T_{CMB}$ producing redshifted 21-cm emission. As shown by Eq. (1), the brightness temperature here depends critically upon the neutral fraction which diminishes as the IGM becomes further ionized. This emission occurs at ν ≈ 100-200 MHz near the end of the Epoch of Reionization (z<12). Ultimately, reionization kills off this redshifted 21-cm signal. Observations of the global HI emission at these frequencies can provide important constraints on the formation and nature of the first black holes.

## OBSERVATIONAL ROADMAP TO THE FIRST FEEDBACK

Highly redshifted 21-cm signals from the Dark Ages and the Epoch of Reionization (EoR) have the potential of producing a very rich cosmological database rivaling that of the CMB. Constraints on the properties of very high-z galaxies and their effects on the surrounding IGM, Dark Energy, Dark Matter Decay, and the nature of the field that drove the Universe during Inflation may emerge from this database [1]. Furthermore, the database will be three dimensional, permitting us to probe the evolution of HI and feedback via redshifted 21-cm tomography.

Efforts are now underway to build the first generation of EoR experiments on the ground to probe the low redshift, higher frequency portion of the spectrum in Fig. 1. This includes EDGES [5], a single broadband antenna with a high dynamic range radio spectrometer currently operating at 100-200 MHz designed to search for a transition in the ionization state of the Universe. Challenges include human interference, the ionosphere, and galactic foregrounds.

In addition, there are three ground-based arrays under construction designed to measure the evolution of the power spectrum of the HI signals on the sky. These include the Murchison Wide-Field Array (MWA) [6] and the Precision Array to Probe the EoR (PAPER) [7] both in western Australia, and the Low Frequency Array (LOFAR) in Europe [8]. This next decade will reveal if these arrays can accurately remove foregrounds at the levels needed to see the redshifted 21-cm EoR signal. This could lead to the development of a low frequency Square-Kilometer Array (SKA) [9].

The farside of the Moon has significant advantages since it effectively lacks an ionosphere, has a demonstrated radio-quiet environment free of interference [10], and is shielded from powerful solar emissions during the lunar night. However, Galactic foregrounds, which become stronger at lower frequencies ($T_{sky} \sim \nu^{-2.6}$), remain a substantial obstacle.

The ground-based low frequency telescopes have the potential to detect the higher frequency, lower redshift HI components of the EoR. However, to probe the beginning of the EoR and into the Dark Ages, the lunar farside is likely the only location in the inner solar system from which such observations will be successful.

A single dipole antenna aboard a spacecraft in lunar orbit or on the lunar surface operating in the frequency regime of ≈ 25-75 MHz has much potential. A stable broadband spectrometer with carefully controlled noise on the spacecraft bus will be required to carry out high dynamic range observations. Innovative, robust algorithms for removal of the significant foregrounds will be required. But, such observations have the potential to test models of the expected average all-sky absorption signals as illustrated in Fig. 1. These data would be the first to sample the formation of the first stars and black holes at z ≈ 10-20, thus providing concrete measurements on how the universe was reionized during this first generation of feedback.

A next step will be a modest-size interferometer of ~100 elements deployed on the lunar farside. Such an array could verify and extend to lower frequencies the results on the EoR from ground-based low frequency telescopes. Such an interferometer could be deployed robotically with astronaut supervision. This precursor telescope would then evolve into a more capable ~$10^4$ element array for imaging tomography of the beginning of the EoR and the Dark Ages [11]. The deployment of such a Lunar Radio Array will be facilitated by NASA's planned Ares V heavy-launch vehicle.

## ACKNOWLEDGMENTS

The LUNAR consortium, headquartered at the University of Colorado, is funded by the NASA Lunar Science Institute to investigate concepts for astrophysical observatories on the Moon. I wish to thank the NLSI for its support and my collaborators in this effort including J. Lazio, J. Hewitt, C. Carilli, J. Kasper, R. MacDowall, S. Furlanetto, J. Bowman, B. O'Shea, and E. Hallman.

## REFERENCES

1. Furlanetto, S.R., Oh, S.P., & Briggs, F.H. 2006, Phys. Reports, 433, 181.
2. Pritchard, J.R. & Loeb, A. 2008, Phys. Rev. D., 78, 103511.
3. Zaldarriaga, M., Furlanetto, S.R., & Hernquist, L. 2004, ApJ, 608, 622.
4. P. Madau, A. Meiksin, M. J. Rees, Astrophys. J. 475 (1997) 429.
5. Bowman, J.D., Rogers, A.E.E., & Hewitt, J.N. 2008, ApJ, 676, 1.
6. MWA: http://www.mwatelescope.org/.
7. PAPER: http://astro.berkeley.edu/%7Edbacker/eor/.
8. LOFAR: http://www.lofar.org/.
9. SKA: http://www.skatelescope.org/.
10. Alexander, J.K., & Kaiser, M.L. 1976, JGR, 81, 5948.
11. Lazio, J.T. 2009, SPIE Proceedings, in press.